\documentclass[prl,twocolumn,showpacs,amsmath,amssymb]{revtex4-1}
\usepackage{subfigure}
\usepackage{graphicx}
\usepackage{graphicx}
\usepackage{array}
\usepackage{xcolor}

\usepackage{lineno}

\usepackage{LamcNPi-defs}

\usepackage{hyperref}
\hypersetup
{
colorlinks=true,
linkcolor=blue,
filecolor=blue,
urlcolor=blue,
citecolor=black,
}

\begin{document}
\normalsize
\parskip=5pt plus 1pt minus 1pt


\title{Observation of the Singly Cabibbo-Suppressed Decay \LamCSigKPi}

\author{
	\begin{small}
	   \begin{center}
           M.~Ablikim$^{1}$, M.~N.~Achasov$^{5,b}$, P.~Adlarson$^{74}$, X.~C.~Ai$^{80}$, R.~Aliberti$^{35}$, A.~Amoroso$^{73A,73C}$, M.~R.~An$^{39}$, Q.~An$^{70,57}$, Y.~Bai$^{56}$, O.~Bakina$^{36}$
% ...
\\
            \vspace{0.2cm}
            (BESIII Collaboration)\\
            \vspace{0.2cm} {\it
$^{1}$ Institute of High Energy Physics, Beijing 100049, People's Republic of China\\
% ...
$^{5}$ Budker Institute of Nuclear Physics SB RAS (BINP), Novosibirsk 630090, Russia\\
% ...
$^{35}$ Johannes Gutenberg University of Mainz, Johann-Joachim-Becher-Weg 45, D-55099 Mainz, Germany\\
$^{36}$ Joint Institute for Nuclear Research, 141980 Dubna, Moscow region, Russia\\
% ...
$^{39}$ Liaoning Normal University, Dalian 116029, People's Republic of China\\
% ...
$^{56}$ Southeast University, Nanjing 211100, People's Republic of China\\
$^{57}$ State Key Laboratory of Particle Detection and Electronics, Beijing 100049, Hefei 230026, People's Republic of China\\
% ...
$^{70}$ University of Science and Technology of China, Hefei 230026, People's Republic of China\\
% ...
$^{73}$ University of Turin and INFN, (A)University of Turin, I-10125, Turin, Italy; (B)University of Eastern Piedmont, I-15121, Alessandria, Italy; (C)INFN, I-10125, Turin, Italy\\
$^{74}$ Uppsala University, Box 516, SE-75120 Uppsala, Sweden\\
% ...
$^{80}$ Zhengzhou University, Zhengzhou 450001, People's Republic of China\\
\vspace{0.2cm}
% ...
$^{b}$ Also at the Novosibirsk State University, Novosibirsk, 630090, Russia\\
% ...
}
\end{center}
\vspace{0.4cm}
\end{small}
}

\date{May 9, 2023}

\begin{abstract}

The singly Cabibbo-suppressed decay \LamCSigKPi{} is observed for the first time with a 
statistical significance of $5.4\sigma$ by using 4.5~\ifb{} of \ee{} collision data collected at 
center-of-mass~energies between 4.600 and 4.699~GeV with the BESIII detector at BEPCII. 
The absolute branching fraction of \LamCSigKPi{} is measured to be 
$(3.8\pm1.2_{\rm stat}\pm0.2_{\rm syst})\times 10^{-4}$ in a model-independent approach. 
This is the first observation of a Cabibbo-suppressed \LamC{} decay involving $\Sigma^-$ in the final state. 
The ratio of branching fractions between $\LamCSigKPi$ and the Cabibbo-favored decay 
$\LamC{}\to \Sigma^- \pi^+\pi^+$ is observed to be $(0.4 \pm 0.1)s_{c}^{2}$,
where  $s_{c} \equiv \sin\theta_c = 0.2248$ with $\theta_c$ the Cabibbo mixing angle.

\end{abstract}

\maketitle

Understanding the nonfactorization contribution is critically challenging for advancing our knowledge of the hadronic weak decays of charmed baryons,
as $W$-exchange and inner $W$-emission are no longer subject to helicity and color suppression~\cite{PhysRevD.44.2799}. The evaluation of nonfactorizable terms is far more difficult than that of factorizable ones and thus the constraints with experimental results are essential. Studies on the charmed baryon decays in experiment have promoted the understanding on the mechanism of charmed baryon decays. Nevertheless, there are still some puzzles needed to be understood. Fo example, why does the breaking effects arising from $m_{s}\gg m_{u,d}$ under SU(3) flavor symmetry significantly exist in the decays of $\Xi_{c}^0$~\cite{GENG2018593}, but much smaller in $\LamC$ decays~\cite{PDG:2022,BESIII:lambdak}?
Furthermore, the discrepancy between the experimental results and the theoretical prediction for the branching fraction (BF) of the decay $\LamC\to p\pi^0$ may indicates the significant contribution of a nonfactorization component and the interference between nonfactorization and factorization contributions~\cite{Cheng:2018,Geng:2018,Geng:2019}. Therefore, correctly estimate the nonfactorization contribution is still one of core tasks in the charm baryon physics.

\begin{figure}[!htp]
    \begin{center}
            \includegraphics[width=0.45\textwidth,height=0.15\textheight]{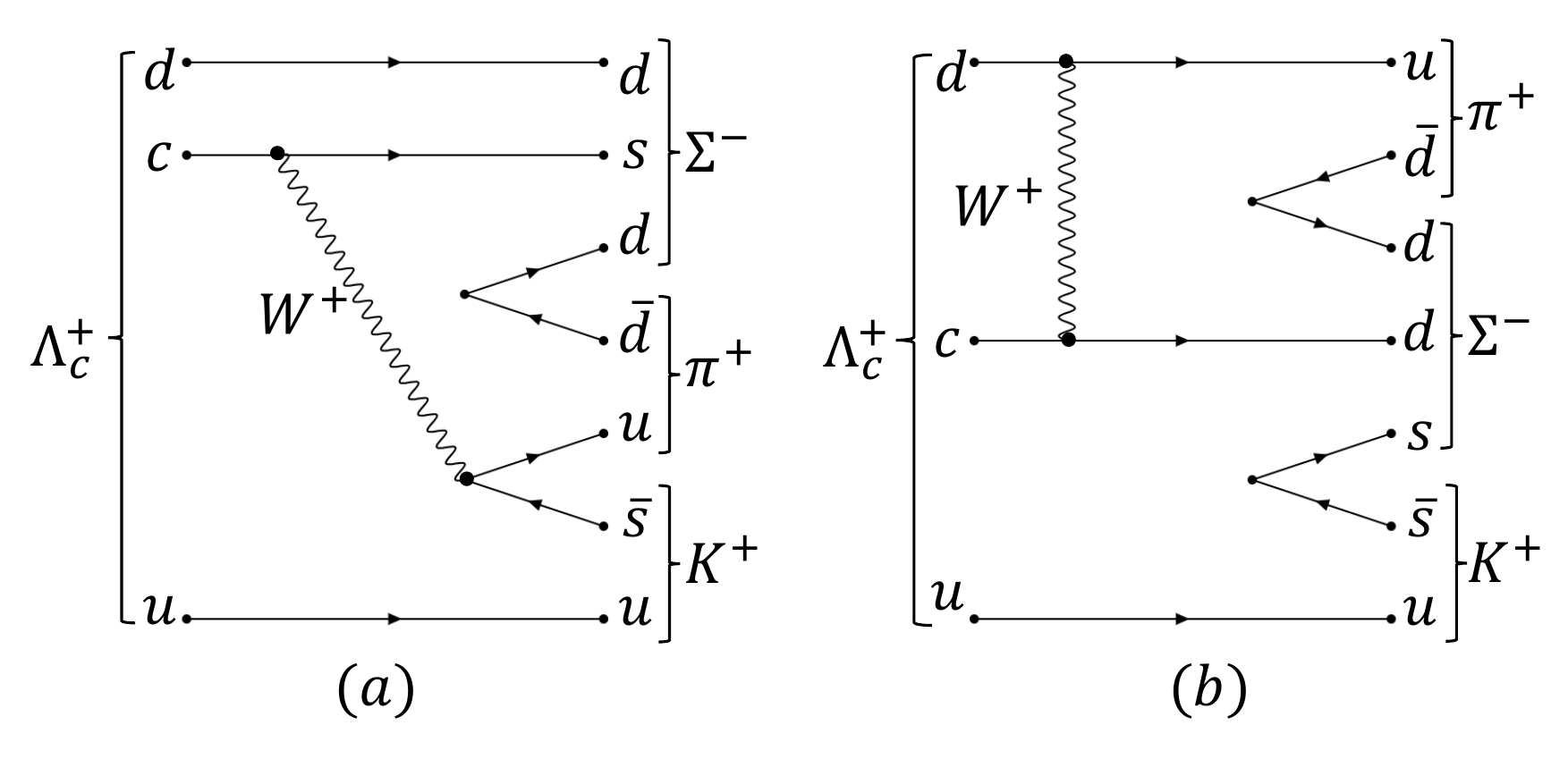}
	\end{center}
    \caption{Feynman diagrams for the CS decay \LamCSigKPi{}: (a) internal $W$-emission, (b) $W$-exchange.
       }
        \label{fig:feynman}
\end{figure}

Extensive studies on Cabibbo suppressed (CS) decays of charmed baryon  in both experiment and theory have been conducted 
for two-body decays~\cite{Cheng:2015Front,Cheng:2021qpd}, because both the factorization and nonfactorization contributions are involved. 
But majority of them could not be well described by phenomenological models~\cite{BESIII:npi,Cheng:2018,Cheng:2018,Geng:2018,Geng:2019,Sharma:1997,Uppal:1994,Lu:2016,Zhao:2020,Cheng:2020}. This indicates the current description of the nonfactorization 
contribution is still not fully reliable. More experimental information is desirable, especially for the decays with hyperons, because the experimental results of CS processes with a hyperon in the final state are still limited. 
Up till now, data for three body CS decays exist only for the $\Sigma^+$. The decay $\LamC\to \Sigma^-K^+\pi^+$ is the simplest singly CS process with a $\Sigma^-$ directly
in the final state, where the $W$-exchange and inner $W$-emission diagrams are expected to play the dominant role, as shown in Fig.~\ref{fig:feynman}. 
Therefore, the observation of the  CS process \LamCSigKPi{} 
and the comparison with the Cabbibo favored (CF) decay $\LamC\to \Sigma^-\pi^+\pi^+$ 
will open a new window for probing  SU(3)$_{F}$ breaking effects and the nonfactorization contribution in $\LamC$ decays.

According to recent constraints provided by the reported BF of the inclusive decay $\LamC\to n+X$~\cite{BESIII:nx}, 
there is a large room to probe experimentally for decays with a neutron in the final state,
including the $\LamC$ decays to $\Sigma^-$, as $\Sigma^-\to n\pi^-$ almost saturates the BF. 
At present, two-body CS processes involving the lighter baryons ($p$~\cite{PhysRevLett.117.232002,JHEP03(2018)043}, 
$n$~\cite{BESIII:npi}, $\Lambda$~\cite{PhysRevD.75.052002,BESIII:lambdak}) 
or $\Sigma^{+/0}$ hyperon ~\cite{PhysRevD.75.052002,ABE200233})
from $\LamC$ decays have already been confirmed and studied extensively in experiment. However, no CS decays with a $\Sigma^-$ have been observed.

Processes with $\Sigma^-$ can be investigated by reconstructing the neutron signal through its missing energy under energy-momentum conservation at BESIII.  
Starting from threshold for $\Lambda_c^+$ pair production at 4.600 GeV, in this Letter, the first observation of the singly CS decay \LamCSigKPi{} is reported using 
4.5~\ifb{} of \ee{} collision data collected with the BESIII detector at seven center-of-mass (c.m.)~energies between 4.600 and 4.699~GeV~\cite{BAIQIANKE}. Throughout this Letter, charge-conjugate modes are implicitly included. 

Details about the design and performance of the BESIII detector can be found in Ref.~\cite{Ablikim:2009aa}.
Simulated samples are produced with a Geant4-based~\cite{Agostinelli:2002hh} Monte Carlo (MC) toolkit,
which includes the geometric description~\cite{Kaixuan:2022} of the BESIII detector. Signal MC samples of 
$\ee\to\LamC\ALamC$ with \ALamC{} decaying into ten hadronic modes and 
$\LamC$ to $\Sigma^-K^+\pi^+$, $\Xi^-K^+\pi^+$, and $\Xi^{*0}K^+$ are used to determine the detection efficiencies,
where the intermediate states are required to be $\Sigma^-\to n\pi^-$ and $\Xi^{\ast 0}\to \Xi^-\pi^+$ with $\Xi^-$ subsequently decaying through any allowed process.
The ten hadronic decay modes are presented in Table~\ref{tab:Effi-Sigma}.
These samples are generated for the individual c.m.~energy by the generator {\sc kkmc}~\cite{Jadach:2000ir} incorporating initial-state radiation effects and the beam-energy spread.
The inclusive MC samples, consisting of open-charm states, radiative return to charmonium (like) $\psi$ states, 
and continuum processes $e^{+}e^{-}\rightarrow q\bar{q}$ ($q=u,d,s$), are generated to survey potential backgrounds. 
Particle decays are modeled with {\sc evtgen}~\cite{Lange:2001uf, Ping:2008zz} 
using BFs taken from the Particle Data Group (PDG)~\cite{PDG:2022}, 
when available, or otherwise estimated with {\sc lundcharm}~\cite{Chen:2000tv,PhysRevLett.31.061301}. Final-state radiation from charged
final-state particles is incorporated using {\sc photos}~\cite{Richter-Was:1992hxq}.

The double-tag (DT) approach is employed to measure the absolute BF of 
$\LamCSigKPi$. A data sample of \ALamC{} baryons, referred to as the single-tag (ST) sample, 
is reconstructed with
ten exclusive hadronic decay modes, as the aforementioned and listed in Table~\ref{tab:Effi-Sigma}. 
The procedure of selecting the ST $\ALamC$ baryon decays is described in 
Refs.~\cite{BESIII:npi,PhysRevD.106.072002,PhysRevD.106.072008}, 
where $105249 \pm 386$ ST events are reconstructed in data. The fit curves for the beam-constrained mass $M_{BC}$ of ST modes and their yields are summarized in Supplementary~\cite{Supp:2023}. 
Those events in which the signal decay $\LamCSigKPi$ is reconstructed in the system recoiling 
against the \ALamC{} candidates of the ST sample are denoted as DT candidates. 

The decay \LamCSigKPi{} with $\Sigma^- \to n\pi^-$ is searched for among the remaining tracks
recoiling against the ST \ALamC{} candidates. Particle identification (PID) is implemented by combining measurements of the ionization energy loss in the helium-based multilayer drift chamber (MDC) ($d$E/d$x$) and the flight time in the time-of-flight system. Only three charged tracks, detected in the MDC and reconstructed within a polar angle ($\theta$) range of $|\!\cos\theta|<0.93$, set by the drift chamber acceptance, are allowed for a DT signal candidate event, where $\theta$ is defined with respect to the $z$ axis, which is the symmetry axis of the MDC.   Two of the charged tracks, whose distances of closest approach to the interaction point (IP) must be less than 10~cm along the $z$ axis ($|V_z| < 10$ cm) and less than 1~cm in the transverse plane ($|V_r| < 1$~cm), are assigned to be $K^+$ and $\pi^+$, according to the PID probability. A vertex fit is performed to the $K^+$ and $\pi^+$ candidates, and the momenta updated by the fit are used in the subsequent analysis.
A third track, identified as a $\pi^-$, is assigned to originate from the $\Sigma^-$
decay if its distance of the closest approach to the IP is within $\pm 20$~cm along the $z$ axis ($|V_z| < 20$~cm). To suppress background events containing other long-lived particles in the final state, the candidate events are further required to have no extra charged tracks with $|\!\cos\theta|<0.93$, $|V_r| < 1$~cm and $|V_z| < 20$~cm.

The neutron signal could be observed in $M_{\rm rec}(B^0)$, where the recoiling mass $M_{\rm rec}(B^0)$ is calculated as:
\begin{equation} \label{eq:mrec2}
\begin{aligned}
\left[M_{\rm rec}(B^0)\right]^2 = \left[\Ebeam - \sum_{i} E_i\right]^2/c^4 
    \\ - \left| \rho\cdot\vec{p}_{0} - \sum_{i}\vec{p}_i \right|^2/c^2 .
\end{aligned}
\end{equation}
Here $E_i$ and $\vec{p}_i$ represent the energy and momentum, respectively, of particle $i$ ($K^+$, $\pi^+$ or $\pi^-$), 
$\rho = \sqrt{\Ebeam^2/c^2 - m_{\LamC}^2 c^2}$,
and $\hat{p}_{0} = - \pALC/|\pALC|$ is the unit direction opposite to the ST \ALamC{},
where  $m_{\LamC}$ is the known $\LamC$ mass~\cite{PDG:2022}. The $\Sigma^-$ signal reconstructed through $M_{\rm rec}(H^-)$, which is as defined in Equation~\ref{eq:mrec2}, with the subscript 
$i$ now representing the $K^+$ and $\pi^+$ particles.
To suppress the continuum hadron background (denoted as $q \bar q$ hereafter), 
the recoiling mass against the ST \ALamC{} in the center-of-mass frame, defined as
\begin{equation} \label{eq:mrec1}
\begin{aligned}
   M_{\rm rec}(\ALamC) = \sqrt{(2\Ebeam - E_{\ALamC})^2/c^4 - | \vec{p}_{\ALamC}|^2/c^2},
\end{aligned}
\end{equation}
is required to fall inside the range (2.275, 2.310)~GeV/$c^2$, 
where $\Ebeam$ is the beam energy, and $E_{\ALamC}$ and $\vec{p}_{\ALamC}$ are the energy and momentum of the ST \ALamC{}, respectively. 
To remove the peaking background due to the process \LamCSigplusKPi{}, we exclude  events
with $M_{\rm rec}(H^+) \in (1.15,1.24)$~GeV/c$^2$, where the recoiling mass $M_{\rm rec}(H^+)$ 
is as defined in Equation~\ref{eq:mrec2}, with the subscript $i$ now representing the $K^+$ and $\pi^-$ particles.
Additionally, to suppress the potential background from $\LamC \to n \Ks K^+$ decays, 
events satisfying $M(\pi^+\pi^-) \in (0.48,0.52)$~$\mathrm{GeV}/c^2$ are vetoed where 
$M(\pi^+\pi^-)$ is the invariant mass of the $\pi^+\pi^-$ pair. 
The requirement $M_{\rm rec}(H^-) > 1.15$~$\mathrm{GeV}/c^2$ is imposed to suppress
backgrounds due to the $q \bar q$ and non-signal \LamC\ALamC{} processes. 

The two-dimensional (2D) distribution of $M_{\rm rec}(H^-)$ and $M_{\rm rec}(B^0)$ is shown in Fig.~\ref{fig:2D}, where the events containing both a $\Sigma^-$ and a neutron
are clustered close to the left-bottom corner, indicating the existence of the singly CS decay 
$\LamCSigKPi$ with $\Sigma^-\to n\pi^-$.
A large number of events containing a $\Lambda$ and $\Xi^-$ appear in the central region
of the 2D distribution, which originate from the CF decay $\LamC\to \Xi^- K^+\pi^-$ with 
$\Xi^-\to\Lambda\pi^-$ and $\Lambda$ decaying into neutral particles. These resonances
are also observed in the projected distributions of $M_{\rm rec}(B^0)$ and 
$M_{\rm rec}(H^-)$, as shown in Fig.~\ref{fig:Kpipi} and \ref{fig:Kpi}, respectively.
Furthermore, by selecting the events in the $\Xi^-$ signal region
$M_{\rm rec}(H^-) \in (1.294,1.340)$ GeV/$c^2$, the $\Xi^{\ast 0}$ signal
is observed, originating from the process $\LamC\to \Xi^{\ast 0} K^+$ with 
$\Xi^{\ast 0}\to \Xi^-\pi^+$, in the distribution of $M_{\rm rec}(K^+)$  as shown in Fig.~\ref{fig:XisK}. 
Here the variable $M_{\rm rec}(K^+)$ is also determined according to
Equation~\ref{eq:mrec2}, with the subscript $i$ labeling only the $K^+$ particle.

\begin{figure}[!htp]
    \begin{center}
            \subfigure{\includegraphics[width=0.2\textwidth,height=0.11\textheight, trim=5 0 0 0, clip]{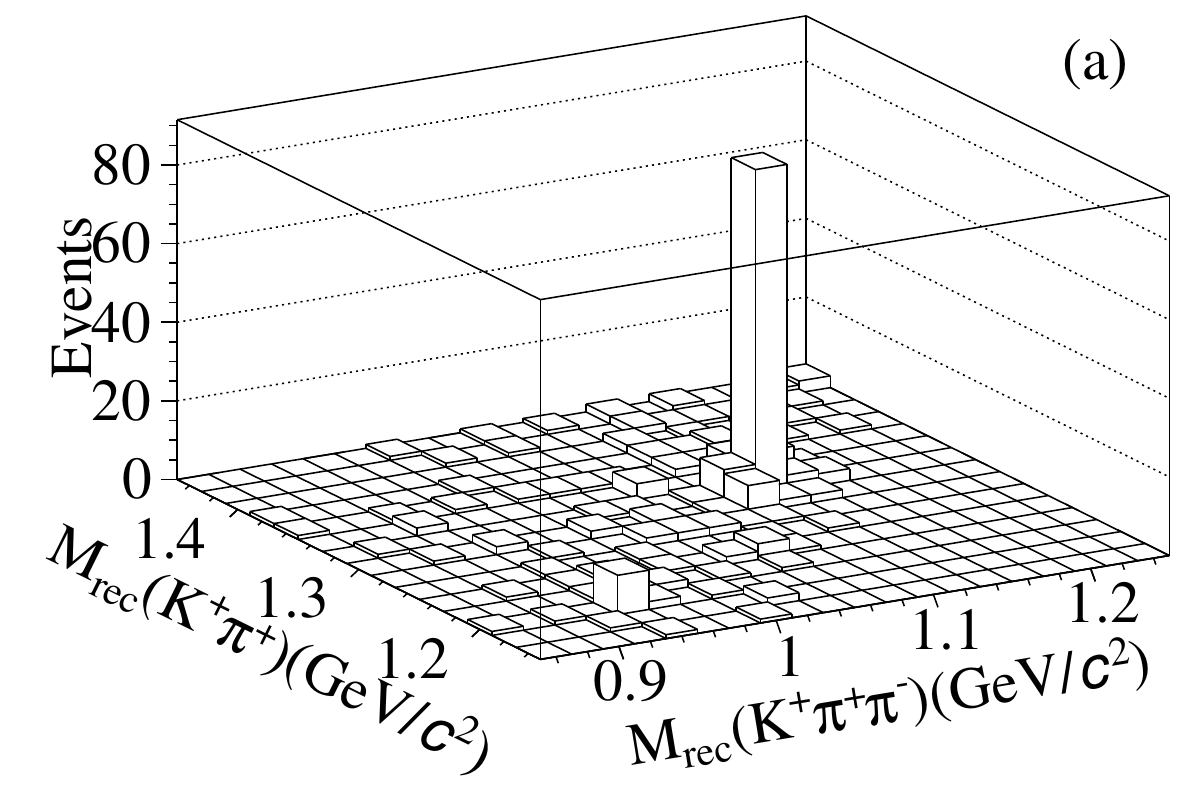}~\label{fig:2D}} 
            \subfigure{\includegraphics[width=0.2\textwidth,height=0.11\textheight, trim=5 0 0 0, clip]{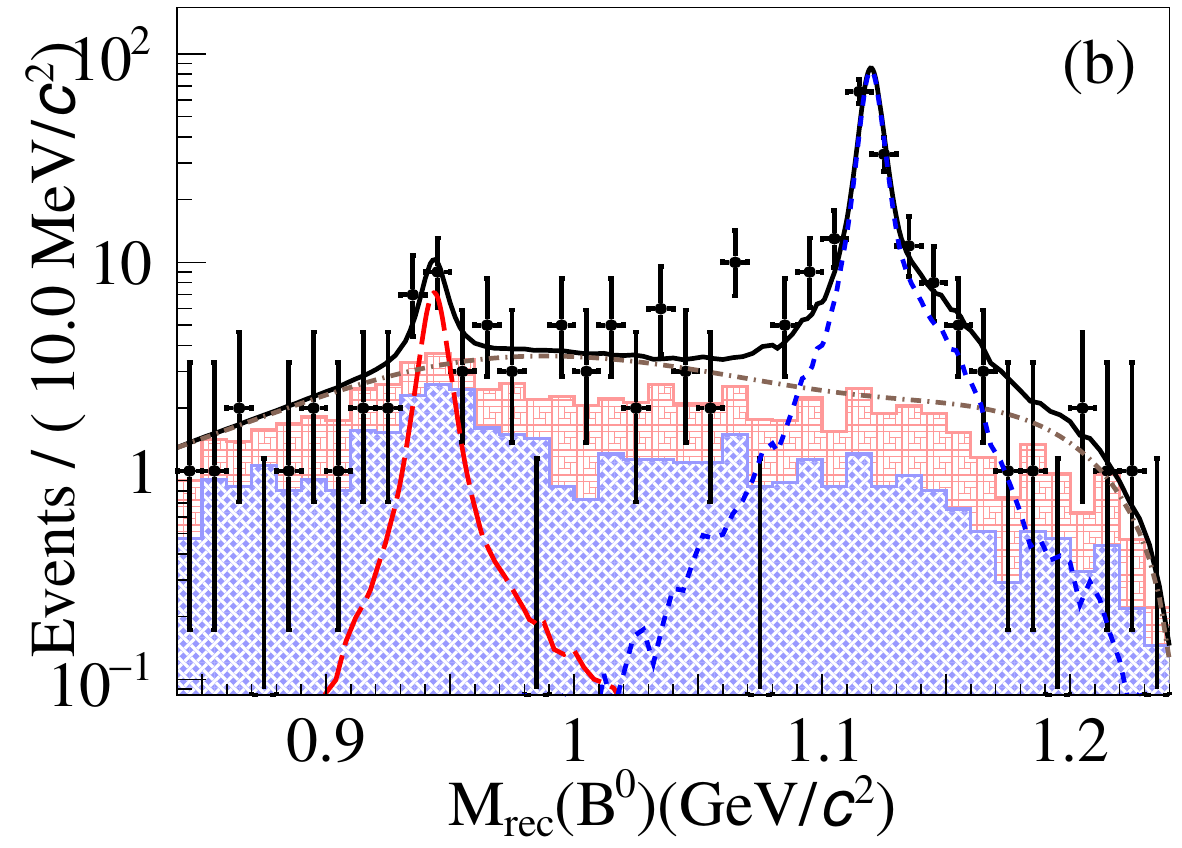}~\label{fig:Kpipi}}  
            \subfigure{\includegraphics[width=0.2\textwidth,height=0.11\textheight, trim=5 0 0 0, clip]{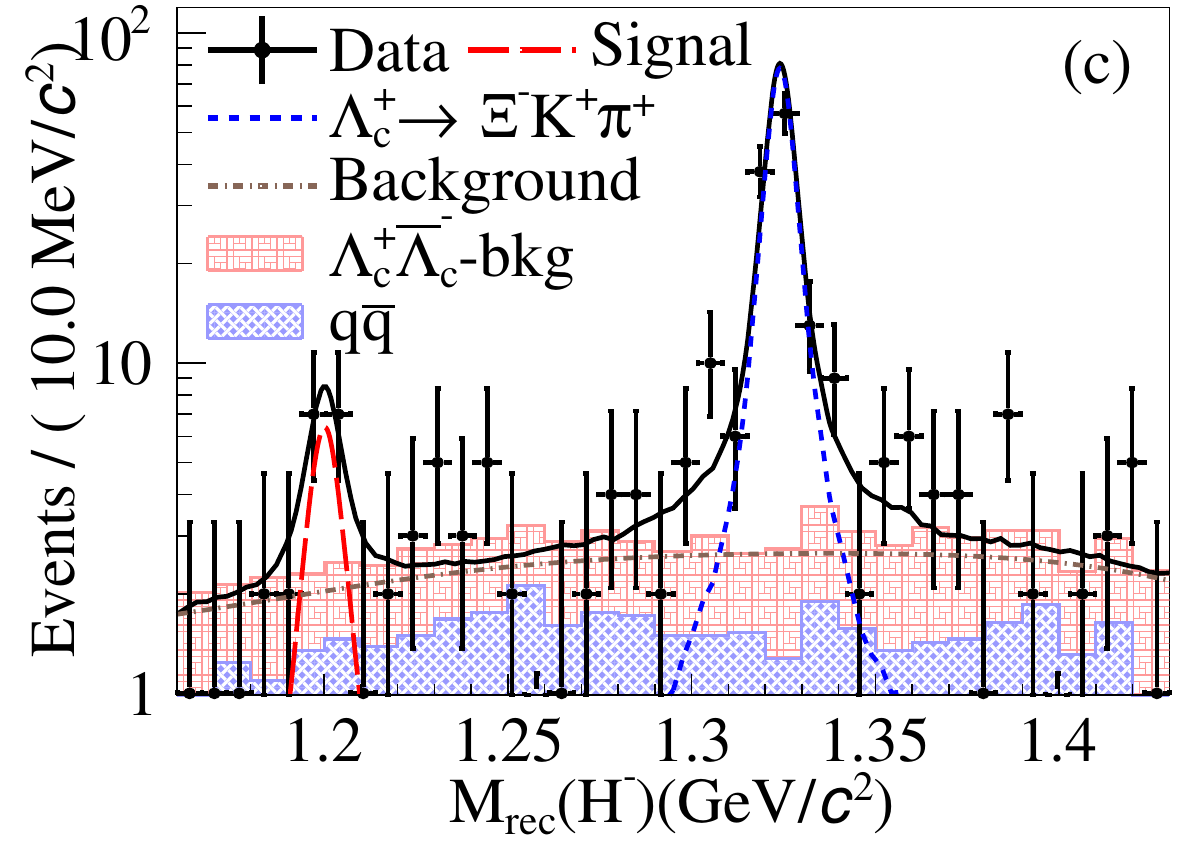}~\label{fig:Kpi}}
            \subfigure{\includegraphics[width=0.2\textwidth,height=0.11\textheight, trim=5 0 0 0, clip]{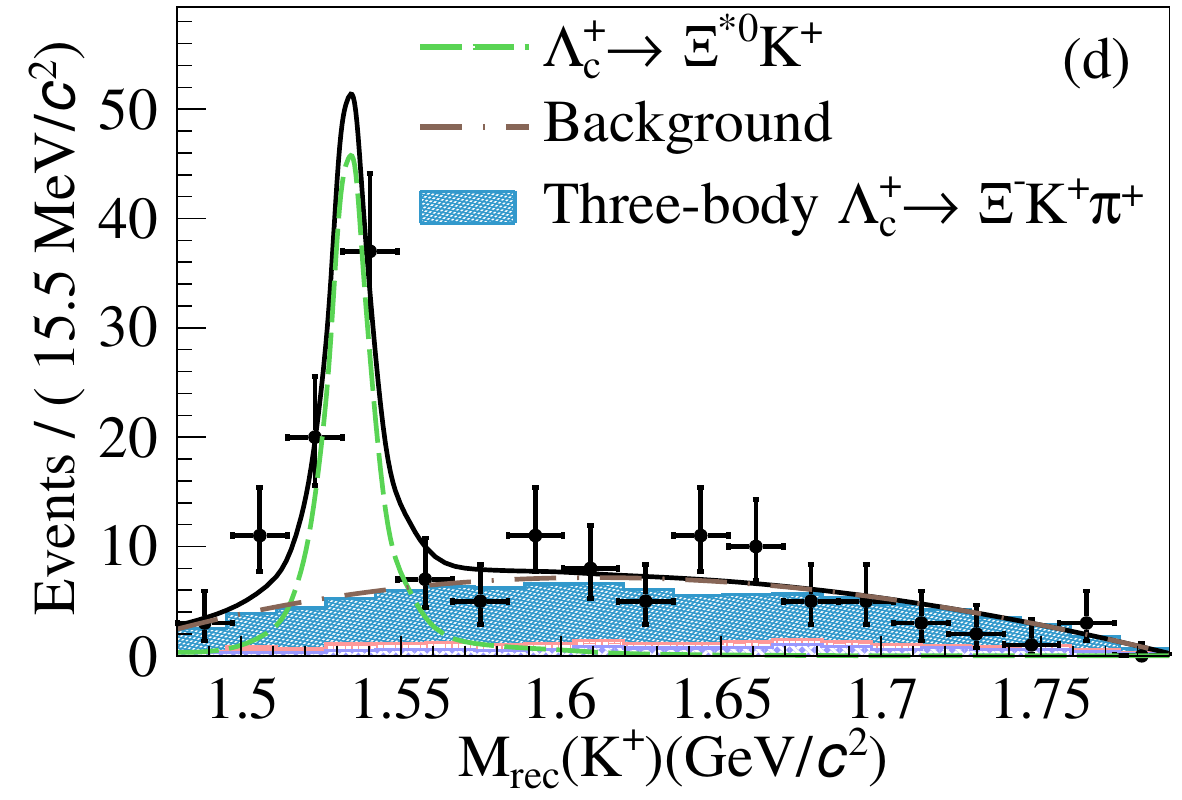}~\label{fig:XisK}}
	\end{center}
    \caption{The 2D distribution of $M_{\mathrm{rec}}(K^+\pi^+)$ versus $M_{\mathrm{rec}}(K^+\pi^+\pi^-)$~(a), the distributions of 
    $M_{\mathrm{rec}}(K^+\pi^+\pi^-)$~(b), $M_{\mathrm{rec}}(K^+\pi^+)$~(c),  and $M_{\mathrm{rec}}(K^+)$~(d) of the accepted DT candidate events from data for all energy points. The black points with error bars are data. The curves represent the fit results, including the signal
   and background components, respectively.
       }
        \label{fig:fit}
\end{figure}

Potential backgrounds are classified into two categories: $q\bar{q}$ processes, and $\ee\to\LamC\ALamC$ events excluding signal contributions of 
$\LamCSigKPi$, $\Xi^-K^+\pi^+$, and $\Xi^{*0}K^+$ (referred to as  $\LamC\ALamC$ background hereafter). 
The $q\bar{q}$ and $\LamC\ALamC$ backgrounds are investigated with the inclusive MC samples with an integrated luminosity 40 times higher than that of data,
and they are normalized to the same integrated luminosity as the data. No peaking background is observed in these samples. In Fig.~\ref{fig:Kpipi} and \ref{fig:Kpi}, the components of $q \bar q$ and $\LamC \bar{\Lambda}_{c}^{-}$ backgrounds are described with the inclusive MC samples that are normalized with the scale factor 0.034. The scale factor is obtained by comparing the number of events between data and inclusive MC samples in the sideband region $M_{BC} \in (2.10, 2.25)$  GeV/$c^2$ of ST $\bar{\Lambda}_{c}^{-}$.  

The signal yields ($N_{\rm obs}$) of the  $\LamCSigKPi$ and $\Xi^- K^+\pi^+$ decays are obtained 
by performing an unbinned maximum-likelihood fit 
to the 2D distribution of $M_{\rm rec}(H^-)$ and $M_{\rm rec}(B^0)$,
where the 2D signal shapes are modeled by the simulated shapes for the two decays, respectively, 
convolved with the same Gaussian function 
accounting for the resolution difference between data and MC simulation.
The 2D shape of $q\bar{q}$ and $\LamC\ALamC$ backgrounds is modeled with the product of two third-order Chebyshev polynomial functions and a Student distribution\cite{student.t,PhysRevD.89.091103,besiiicollaboration2023evidence,besiiicollaboration2023amplitude} that is used to describe the dispersion of the backgrounds in the diagonal direction. Details of the background functions and the validation are given in the Supplementary~\cite{Supp:2023}. 
Additionally, a fit to the distribution of $M_{\rm rec}(K^+)$ is performed simultaneously 
to determine the yield of the decay $\LamCXistarK$, where its shape is also described
from the simulation convolved with an individual Gaussian function.
Here, the $q\bar{q}$ and $\LamC\ALamC$ backgrounds
are described individually by two third-order Chebyshev polynomials,
whose shape parameters are obtained from fits to the corresponding inclusive 
MC samples, and whose magnitudes are determined from the fit to the data.
In addition, the non-resonant three-body decay $\LamCXiKPi$ 
has a smooth distribution in $M_{\rm rec}(K^+)$, and is also modeled by 
a third-order Chebyshev polynomial with its shape parameters obtained from the MC simulation, and its magnitude determined from the fit to data.  
The yield of $\LamCXiKPi$ decays in the 2D fit is constrained to be equal to the sum
of those of the decay $\LamCXistarK$ and the three-body decay $\LamCXiKPi$ in the fit to the distribution 
of $M_{\rm rec}(K^+)$. The yield of $q\bar{q}$ and $\LamC\ALamC$ backgrounds in the region 
$M_{\rm rec}(H^-) \in (1.294,1.340)$~$\rm{GeV}/c^2$ is fixed to the numbers obtained from the fit to the distribution of $M_{\rm rec}(K^+)$.
The resultant fit is depicted in Fig.~\ref{fig:fit}, and the signal yields 
are determined to be $12 \pm 4$,  $128 \pm 13$, and $54 \pm 8$ for $\LamCSigKPi$, $\Xi^- K^+ \pi^+$ 
and $\Xi^{*0} K^+$, respectively, where the uncertainties are statistical. 
The statistical significance of  the $\LamCSigKPi$ signal is $5.4\sigma$, which is calculated from the 
change of the likelihood values between fits with and without the signal component, accounting for the change in the number of degrees of freedom and taking into account both statistical and systematic uncertainties. The details of how to take the systematic uncertainty into account are shown in the Supplementary~\cite{Supp:2023}.

\begin{table*}[!htbp]
\centering
  \begin{center}
  \caption{The DT detection efficiencies (\%) for $\LamCSigKPi$/$\LamCXiKPi$/$\LamCXistarK$ 
    for each ST mode at c.m.~energies between $\sqrt{s} = $ 4.600 and  4.699~GeV.}
  \renewcommand\arraystretch{1.2}\footnotesize
\begin{tabular}{ l r @{$/$} c @{$/$} l r @{$/$} c @{$/$} l r @{$/$} c @{$/$} l r @{$/$} c @{$/$} l r @{$/$} c @{$/$} l r @{$/$} c @{$/$} l r @{$/$} c @{$/$} l}
      \hline
      \hline

           channel / $\sqrt s$ (GeV)& \multicolumn{3}{c}{4.600} & \multicolumn{3}{c}{4.612} & \multicolumn{3}{c}{4.628} & \multicolumn{3}{c}{4.641} & \multicolumn{3}{c}{4.661} &  \multicolumn{3}{c}{4.682} & \multicolumn{3}{c}{4.699}\\
      \hline
	     $\apkpi$                         & 16.5 & 8.5 & 8.2   &  15.6 & 8.0 & 7.8 &  15.1 & 7.7 & 7.6 &  15.0 & 7.7 & 7.7 & 14.8 & 7.4 & 7.6 & 14.1 & 7.4 & 7.3 & 14.0 & 7.2 & 7.3 \\
            $\apks$                        & 18.7 & 9.7 & 9.7 &  17.5 & 9.0 & 9.1 & 16.9 & 8.5 & 8.2  &    16.5 & 8.5 & 8.6 & 15.7 & 8.3 & 8.3 & 15.7 & 8.0 & 8.1 & 15.1 & 7.8 & 7.9 \\
            $\apkpi\pi^0$                  & 4.6 & 1.9 & 1.9  &  4.2 & 2.3 & 2.0  & 4.1 & 2.2 & 2.0  &  4.4 & 2.2 & 1.8  & 3.9 & 2.1 & 1.7  & 4.3 & 1.9 & 1.6  & 4.2 & 2.1 & 1.7 \\
            $\apks\pi^0$                   & 6.5 & 3.3 & 3.3  &  6.0 & 3.0 & 2.8  & 5.2 & 2.7 & 2.6  &  5.3 & 2.9 & 2.8 & 5.2 & 2.8 & 2.7  & 5.3 & 2.8 & 2.6  & 5.2 & 2.6 & 2.5 \\
            $\apks\pi^+\pi^-$              & 6.2 & 3.1 & 3.1  &  5.5 & 2.8 & 2.8  & 5.2 & 2.8 & 2.6  &  5.5 & 2.7 & 2.7 & 5.2 & 2.9 & 2.9  & 5.0 & 2.6 & 2.6  & 4.9 & 2.6 & 2.5 \\
            $\bar{\Lambda}\pi^-$           & 14.5 & 7.1 & 7.1 &  12.7 & 7.1 & 7.2 & 12.3 & 6.4 & 6.6 &  12.1 & 6.5 & 6.7 & 12.5 & 6.0 & 6.0 & 11.0 & 6.1 & 6.0 & 11.7 & 5.4 & 5.5 \\
            $\bar{\Lambda}\pi^-\pi^0$      & 5.7 & 2.8 & 2.8  &  5.1 & 2.5 & 2.5  & 4.7 & 2.4 & 2.4  &  4.7 & 2.4 & 2.4 & 4.7 & 2.3 & 1.8  & 4.5 & 2.3 & 1.6  & 4.3 & 2.1 & 1.8 \\
            $\bar{\Lambda}\pi^+\pi^-\pi^-$ & 4.0 & 1.9 & 1.9  &  3.6 & 1.9 & 1.8  & 3.7 & 1.7 & 1.7  &  3.6 & 1.7 & 1.6 & 3.3 & 1.9 & 1.8  & 3.6 & 1.8 & 1.6  & 3.6 & 1.8 & 1.8   \\
            $\bar{\Sigma}^0\pi^-$          & 8.2 & 4.0 & 4.0  &  7.3 & 3.7 & 3.6  & 6.4 & 3.2 & 3.1   &  7.1 & 3.3 & 3.1  & 6.8 & 3.1 & 3.2  & 6.8 & 3.1 & 3.2  & 6.0 & 2.9 & 2.7 \\
            $\bar{\Sigma}^-\pi^+\pi^-$     & 6.6 & 3.3 & 3.3  &  6.3 & 3.1 & 3.0  & 6.1 & 3.1 & 3.0  &  5.4 & 2.8 & 2.9  & 5.5 & 2.9 & 2.6  & 5.5 & 2.7 & 2.6  & 5.5 & 2.6 & 2.6  \\
      \hline\hline
    \end{tabular}
          \label{tab:Effi-Sigma}
  \end{center}
\end{table*}

The branching fractions ($\mathcal{B}$) are determined as 
\begin{equation} \label{eq:br}
  \mathcal{B}=\frac{N_{\mathrm{obs}}} {\sum_{ij} N_{ij}^{\mathrm{ST}}\cdot (\epsilon_{ij}^{\mathrm{DT}}/\epsilon_{ij}^{\mathrm{ST}}) },
\end{equation}
where the subscripts $i$ and $j$ label the ST modes and the data samples at individual c.m.~energies, respectively.
The parameters $N_{ij}^{\mathrm{ST}}$, $\epsilon_{ij}^{\mathrm{ST}}$ and $\epsilon_{ij}^{\mathrm{DT}}$ are the ST yields, and ST and DT efficiencies, respectively.
The detection efficiencies $\epsilon_{ij}^{\mathrm{ST}}$ and $\epsilon_{ij}^{\mathrm{DT}}$ are estimated from  signal MC samples, where the key distributions of the ST modes have been reweighted to agree with those of data. 
Since the decay $\LamCXiKPi$ has two major components, {\it i.e.} 
$\LamCXistarK$ and the non-resonant three-body decay $\LamCXiKPi$, its detection efficiencies
combine the contributions from both these two components. To take into account
potential intermediate-resonance effects, the signal MC sample of $\LamCXiKPi$ 
is reweighted to match the data together with the $\LamCXistarK$ component, and  
the DT efficiencies of the decay $\LamCXiKPi$ are derived. Details of the weights could be found in the Supplementary~\cite{Supp:2023}. 
The ST efficiencies can be found in Supplementary~\cite{Supp:2023},
whereas the DT efficiencies are summarized in Table~\ref{tab:Effi-Sigma}. 
The BFs are determined to be 
$\mathcal{B}(\LamCSigKPi)=(3.8\pm1.2\pm0.2)\times 10^{-4}$,
$\mathcal{B}(\LamCXiKPi)=(7.74\pm0.76\pm0.54)\times 10^{-3}$, and 
$\mathcal{B}(\LamCXistarK)=(5.03\pm0.77\pm0.20)\times 10^{-3}$,
where the first uncertainties are statistical and the second are systematic.

Benefiting from the DT approach, the systematic uncertainties associated with the ST selection efficiency cancel out in the BF measurements. Thus, the systematic uncertainties for this measurements comprise those associated 
with the ST yields, the $K^+$ and $\pi^{\pm}$ tracking and PID efficiencies, the requirement on the number of tracks, the determination of the DT signal yields, the BF of the intermediate-state decays and the statistical uncertainties from the signal MC samples.

The uncertainty in the total ST yields is 0.5\%~\cite{PhysRevD.106.072002,PhysRevD.106.072008}, which arises from the statistical uncertainty and
fitting strategy for extracting these yields. 
The uncertainties associated with the $K^+$ and $\pi^{\pm}$ tracking and PID efficiencies are both assigned to be 1.0\%, from studies performed with control samples of $J/\psi\to K^0_S K^\pm \pi^\mp$, $K^0_S \to \pi^+\pi^-$ ~\cite{PhysRevD.99.012003}  and $J/\psi\to\pi^+\pi^-\pi^0$~\cite{BESIII:jpsi} decays. The uncertainty due to the no extra charged track requirement is 2.2\%, which is assigned from studies of a control sample of $\ee\to\LamC\ALamC$ decays, with $\LamC\to p K^-\pi^+$ and the $\ALamC$ decaying into the ten tagged decay modes. The uncertainties from the determination of the DT yields are 3.7\%, 2.1\% and 2.2\% for the decays $\LamCSigKPi$, $\Xi^- K^+ \pi^+$, and $\Xi^{*0} K^+$, respectively, including those from the modeling of $q\bar{q}$ and $\LamC\ALamC$ backgrounds, which are estimated by considering the uncertainty in the scale-factor for the $q \bar q$ estimation and in the parameters of the Chebyshev polynomial functions and the Student distribution for describing the shape of $q\bar{q}$ and $\LamC\ALamC$ backgrounds. The uncertainties in the quoted BFs of $\Xi^-$ and $\Xi^{*0}$ are both 1.4\% for the decays $\LamCXiKPi$ and $\Xi^{*0} K^+$, respectively. 
The uncertainty arising from the MC modeling for \LamCXiKPi{} is investigated by reweighting the MC distribution to data, and comparing with the results obtained between the original and reweighted samples. The resultant uncertainty in the MC modeling is 5.9\%. The uncertainties associated with the finite size of the signal MC samples are 0.5\%. Assuming that all the sources of bias are uncorrelated, the total uncertainties are then taken to be the quadratic sum of the individual contributions, which are 4.9\%, 6.9\% and 4.0\% for $\LamCSigKPi$, $\Xi^- K^+ \pi^+$, and $\Xi^{*0} K^+$, respectively.

In summary, the singly Cabibbo-suppressed decay $\LamCSigKPi$ is observed for the first time 
with a statistical significance of $5.4\sigma$ by analyzing \ee{} collision data samples 
corresponding to a total integrated luminosity of 4.5~\ifb{} collected at c.m. energies 
between 4.600 and 4.699~GeV with the BESIII detector. 
The BF of $\Lambda^+_c\to \Sigma^-K^+\pi^+$ is 
measured to be $(3.8\pm1.2_{\rm stat}\pm0.2_{\rm syst})\times 10^{-4}$ with a model-independent approach. 
This is the first observation of the CS $\LamC$ decay containing a $\Sigma^-$ in the final state. 
The ratio of BFs between  $\LamCSigKPi$  and the CF decay $\LamC{}\to \Sigma^- \pi^+\pi^+$~\cite{PDG:2022} is observed to be 
$(2.03 \pm 0.73)\%\simeq (0.4 \pm 0.1)s_{c}^{2}$, which is close to the ratio 
$\mathcal{B}(\Xi_{c}^0\to \Xi^- K^+)/\mathcal{B}(\Xi_{c}^0\to \Xi^- \pi^+)$ and deviates significantly from $1.0s_{c}^{2}$, while $1.0s_{c}^{2}$ is also consistent with CS/CF ratio of the isospin partner modes $\mathcal{B}(\LamC\to\Sigma^+ K^+ \pi^-)/\mathcal{B}(\LamC\to\Sigma^+
\pi^+ \pi^-)$.
This result suggests nonfactorization contribution is dominate over the factorization one or large SU(3) flavor symmetry breaking effect in three-body decays involving a $\Sigma^-$ baryon.
A prediction based on SU(3)$_{F}$ symmetry gave $\mathcal{B}^{\rm pred}(\LamCSigKPi)=(3.3\pm2.3)\times 10^{-4}$~\cite{PhysRevD.99.073003},
which has a larger uncertainty than our measurement  due to the limited sample sizes of the  channels used as inputs to the calculation.
Our measurement provides  direct information to 
improve the understanding of the \LamC{} decay mechanisms.  
Meanwhile, the BFs of CF decays $\LamCXiKPi$ and $\LamCXistarK$ 
are measured to be $(7.74\pm0.76_{\rm stat}\pm0.54_{\rm syst})\times 10^{-3}$ and  
$(5.03\pm0.77_{\rm stat}\pm0.20_{\rm syst})\times 10^{-3}$, respectively, 
which are consistent with previous results~\cite{PDG:2022}. The measured $\LamCXiKPi$ is the sum of nonresonant three-body decay and $\LamCXistarK$.

The BESIII collaboration thanks the staff of BEPCII, the IHEP computing center and the supercomputing center of the 
University of Science and Technology of China (USTC) for their strong support.
Authors are grateful to Hai-Yang Cheng, Fanrong Xu and Yu-Kuo Hsiao for enlightening discussions. 
This work is supported in part by National Key R\&D Program of China under Contracts Nos. 2020YFA0406400, 2020YFA0406300; 
National Natural Science Foundation of China (NSFC) under Contracts Nos. 11335008, 11625523, 11635010, 11735014, 11822506, 11835012, 11935015, 11935016, 11935018, 11961141012, 12022510, 12025502, 12035009, 12035013, 12061131003, 12005311, 11805086, 11705192, 11950410506; 
the Fundamental Research Funds for the Central Universities, University of Science and Technology of China, Sun Yat-sen University, Lanzhou University, University of Chinese Academy of Sciences;
100 Talents Program of Sun Yat-sen University;
the Chinese Academy of Sciences (CAS) Large-Scale Scientific Facility Program; Joint Large-Scale Scientific Facility Funds of the NSFC and CAS under Contracts Nos. U1732263, U1832207, U1832103, U2032111; 
CAS Key Research Program of Frontier Sciences under Contract No. QYZDJ-SSW-SLH040; 100 Talents Program of CAS; China Postdoctoral Science Foundation under Contracts Nos. 2019M662152, 2020T130636;
The Institute of Nuclear and Particle Physics (INPAC) and Shanghai Key Laboratory for Particle Physics and Cosmology; ERC under Contract No. 758462; 
European Union Horizon 2020 research and innovation programme under Contract No. Marie Sklodowska-Curie grant agreement No 894790; 
German Research Foundation DFG under Contracts Nos. 443159800, Collaborative Research Center CRC 1044, FOR 2359, FOR 2359, GRK 214; 
Istituto Nazionale di Fisica Nucleare, Italy; Ministry of Development of Turkey under Contract No. DPT2006K-120470; National Science and Technology fund; 
Olle Engkvist Foundation under Contract No. 200-0605; STFC (United Kingdom); 
The Knut and Alice Wallenberg Foundation (Sweden) under Contract No. 2016.0157; The Royal Society, UK under Contracts Nos. DH140054, DH160214; 
The Swedish Research Council; U. S. Department of Energy under Contracts Nos. DE-FG02-05ER41374, DE-SC-0012069.

\end{document}